\documentclass[12pt,preprint,longnamesfirst]{aastex}

\newcommand{\avec}{{\bf a}}

\newcommand{\rvec}{{\bf r}}

\newcommand{\svec}{{\bf s}}

\newcommand{\uvec}{{\bf u}}

\newcommand{\xvec}{{\bf x}}

\newcommand{\thetavec}{\mbox{\boldmath{$\theta$}}}

\newcommand{\meters}{\, {\rm m}}
\newcommand{\seconds}{\, {\rm sec}}
\newcommand{\ksec}{\, {\rm ksec}}
\newcommand{\hours}{\, {\rm hr}}
\newcommand{\cm}{\, {\rm cm}}

\newcommand{\keV}{\, {\rm keV}}

\newcommand{\muK}{\, \mu {\rm K}}

\newcommand{\GHz}{\, {\rm GHz}}
\newcommand{\Mpc}{\, {\rm Mpc}}

\newcommand{\del}{\nabla}
\newcommand{\grad}{\del}
\newcommand{\cross}{\times}

\newcommand{\defeq}{\equiv}

\newcommand{\MSun}{M_{\odot}}

\newcommand{\Label}{\label}

\begin{document}

\title{The Distance to Clusters: Correcting for Asphericity}
\author{David C. Fox}
\affil{Physics Department, Harvard University, Jefferson Physical Laboratories,
Cambridge, MA 02138}
\email{davidcfox@post.harvard.edu}
\and
\author{Ue-Li Pen}
\affil{Canadian Institute for Theoretical Astrophysics, University of Toronto,
McLennan Labs,
60 St. George Street,
 Toronto, ON, M5S 3H8, Canada}
\email{pen@cita.utoronto.ca}


\begin{abstract}

X-ray and Sunyaev-Zel'dovich effect observations can be combined to
measure the distance to clusters of galaxies.  The Hubble constant,
$H_0$, can be inferred from the distance to low-redshift clusters. 
With enough clusters to measure the redshift-distance relation out to a
redshift $z \sim 1$, it may be possible to determine the total matter
density, $\Omega_0$, and the cosmological constant, $\Lambda_0$, as
well.  If the intracluster gas distribution is not spherical, but
elongated by a factor of $Z$ along the line of sight, the inferred
distance is increased by $Z$, and $H_0$ is decreased by the same
factor.  Averaging the inferred value of $H_0$ over a sufficiently
large sample of clusters can reduce any systematic bias due to cluster
shapes, provided the clusters are selected without any preferred
orientation.  Even so, elongation contributes significantly to the
variance in the measured distances and in the inferred value of $H_0$.

With the addition of gravitational lensing observations, it is possible
to infer the three-dimensional shape of an individual cluster, provided
the gas is in hydrostatic equilibrium.  We demonstrate a specific
method for finding the shape and correcting the measured distances to
individual clusters.  To test this method, we apply it to artificial
observations of simple model clusters.  We base the artificial X-ray
observations on the Chandra X-ray Observatory.  For the SZ effect, we
assume modest improvements over current observations at the Owens
Valley Radio Observatory.  We recover the true distances to each of our
clusters without detectable bias, and with statistical errors due to
measurement uncertainties of $4$ to $6 \%$.

\end{abstract}

\keywords{cosmic microwave background --- 
distance scale --- 
galaxies: clusters : general ---
gravitational lensing ---
X-rays: galaxies: clusters}

\section{Introduction}

The discovery of X-ray emission from clusters of galaxies revealed that
clusters are filled with ionized gas.  Not long after this discovery,
\citet{SZ} pointed out that the electrons in this gas should also
scatter photons from the Cosmic Microwave Background (CMB), and that
the thermal motion of the electrons would produce a detectable shift in
the CMB spectrum.  This shift is known as the Sunyaev-Zel'dovich (SZ)
effect.  Together, SZ effect and X-ray observations of the intracluster
gas determine the physical size of the cluster along the line of sight.
 Comparing this scale with the apparent angular size of the cluster
yields a direct measurement of the angular diameter distance, assuming
the cluster is spherical \citep{Cav77, Gunn78, SilkWhite78}.  This
geometric estimate is independent of the usual distance ladder.  Using
cluster distance and redshift measurements for low redshift clusters,
the Hubble constant, $H_0$, can be determined.  Measurements of the
redshift-distance relation at redshifts up to $z \sim 1$ would
constrain the matter density, $\Omega_0$, and and the cosmological
constant, $\Lambda_0$, in units of the critical density
\citep{Carlstrom2001}.

We know from X-ray images that clusters are often not round in
projection, and therefore cannot be spherical \citep[see,
e.g.,][]{Mohr95}.  A number of authors have investigated the effect of
asphericity on the inferred value of $H_0$, either with simulations
\citep{ISS95, RSM97, YIS98} or analytic ellipsoidal models of the gas
\citep{Cooray98, Cooray2000, HB98, Sulkanen99, Puy2000}.  For an
individual cluster, the estimated distance will differ from the actual
distance by a factor of the ratio, $Z$, of the size of the cluster
along the line of sight to its size in the plane of the sky. The usual
solution to this problem has been to suggest averaging the inferred
values of the Hubble constant over a sample of clusters.  Here, we
propose a different approach.  We will demonstrate that it is possible
to infer the three-dimensional shape of an individual cluster, and
recover the true distance.

In its simplest form, the idea behind the distance determination is as
follows.  The SZ effect is caused by Thompson scattering of CMB photons
by electrons.  It is therefore proportional to the column density, $N_e
= n_e \Delta l$, where $n_e$ is the electron number density and $\Delta
l$ is a measure of the path length of a given line of sight through the
cluster.  The X-ray emission is due to Bremsstrahlung from electron-ion
collisions, so the X-ray surface brightness, $\Sigma_X$, is
proportional to $n_e^2 \Delta l$.  Squaring $n_e \Delta l$ and dividing
by $n_e^2 \Delta l$ from the X-ray observations, we can eliminate the
electron density to find the path length.  Finally, by comparing this
path length with the angular extent of the cluster, we can infer the
angular diameter distance, $D$, to the cluster.  For clusters at
redshift $z \ll 1$, we can calculate the Hubble constant, $H_0 = cz /
D$.

We have neglected the fact that both the SZ effect and the X-ray
emissivity depend on the temperature of the electrons as well as their
density.  We describe the method in more detail, taking temperature
into account, in \S\ref{degeneracy}.  However, this simplified picture
suffices to illustrate the effect of cluster shapes on the inferred
distance.  The method assumes that the cluster is spherical, so that
the path length along the line of sight and the width of the cluster
perpendicular to the line of sight are the same. If the actual gas
distribution is not spherical, this assumption can introduce a
systematic error in the inferred distance. If the gas distribution is
elongated by a factor of $Z$ along the line of sight, this will
increase the measured SZ effect and X-ray surface brightness by $Z$,
without changing the apparent angular size of the cluster.  As a
result, the inferred distance will be increased by a factor of $Z$
relative to the true distance, and $H_0 = cz / D$ will be
underestimated by $Z$.

It has generally been assumed that averaging the inferred values of the
Hubble constant over different cluster orientations at random would
eliminate any systematic bias.  In fact, the validity of this
assumption seems to depend on both the intrinsic shapes of clusters
(whether prolate, oblate or triaxial) and on the choice of projected
axis used to establish the angular scale \citep{Cooray98, Cooray2000,
Sulkanen99}, though the remaining systematic errors are relatively
small ($3$ --- $10 \%$) if the semi-major projected axis is used.

In addition, removing the systematic effect of elongation requires
averaging over an unbiased sample of clusters.  If clusters are chosen
based on X-ray surface brightness or SZ effect amplitude (or eliminated
based on non-detections), clusters elongated along the line of sight
will be favored \citep{BHA91}.  Clusters elongated due to recent
mergers may be favored because of their higher X-ray luminosity
\citep{RSM97}.  Some authors \citep{RSM97} have suggested selecting
clusters without evidence of recent mergers, but even this may
introduce its own bias if the merger signatures used are not
independent of orientation.

Finally, even with an unbiased sample, asphericity contributes roughly
$15 \%$ to the rms scatter in the value of $H_0$ from individual
clusters \citep{ISS95, RSM97, Sulkanen99}.  Thus measurements are
required for a large ($\sim 25$) cluster sample.  This may limit more
detailed studies of the redshift-distance relation to estimate
$\Omega_0$ and $\Lambda_0$.  Some evidence \citep{Cooray98} suggests
that a large part of the scatter in measurements of $H_0$ from X-ray
and SZ effect observations is due to variations in cluster shapes.

If we had a method of determining the three-dimensional shape of
individual clusters, we could correct for the effect of elongation on
the distance.  This would enable us to reduce the scatter in the
inferred redshift-distance relation.  In addition to reducing the
number of clusters needed to achieve a given level of precision, this
might also allow us to study other effects (to quantify any variations
in the clumpiness of the intracluster gas from one cluster to another,
for instance).

The degeneracy between the distance, $D$, and the elongation factor,
$Z$, is intrinsic to observations of projected cluster properties. 
Therefore, to measure elongation without prior knowledge of $D$, we
need an additional theoretical constraint.  An obvious choice is
hydrostatic equilibrium.  This assumption in turn introduces a new
unknown, the cluster mass distribution.  To compensate, we must add an
additional observational constraint, the projected mass distribution as
measured by gravitational lensing.  We will prove that the assumption
of hydrostatic equilibrium, together with lensing observations, is
sufficient to break the degeneracy and allow us to recover both the
elongation and the distance.

Such an approach has been suggested by \citet{Zaroubi98, Zaroubi2001}. 
\Citet{Reb2000} has also discussed a method for reconstruction of the
three-dimensional structure of clusters.  The perturbative approach of
\citet{Dore2001} may also be useful in inferring asphericity.
\Citet{Cooray98} has suggested a somewhat different way of breaking the
same degeneracy.  While it also combines X-ray, SZ effect, and lensing
observations, it is based on the assumption of a universal (and
measurable) baryon fraction for clusters, rather than on that of
hydrostatic equilibrium.

We will demonstrate that our method can successfully recover the shape
and distance with high accuracy when applied to simple analytical
models of clusters.  We first test the method on clusters which are
axisymmetric about line of sight, and then on triaxial clusters of
arbitrary orientation.  In the future, we will test its ability to do
the same when applied to simulated clusters, which more closely
resemble real ones, at least in their complexity and asymmetry.  We
should note that one advantage of the traditional method of inferring
$H_0$ from SZ effect and X-ray observations is that it does not depend
on hydrostatic equilibrium.  Deviations from hydrostatic equilibrium in
real clusters may introduce a systematic error in the value of $H_0$
inferred from any method which does rely on this assumption.  It is
essential to apply our method to simulated clusters to attempt to
quantify any such systematic error.

In \S\ref{degeneracy}, we will explain the elongation-distance
degeneracy in more detail.  By using ellipsoidal potentials, we will
prove analytically that the hydrostatic equilibrium assumption and weak
lensing observations are sufficient in principle to break the
degeneracy in the case of a cluster axisymmetric about the line of
sight.  In practice, the actual model we use to infer the elongation
from artificial observations uses ellipsoidal mass distributions.  We
will formulate this method in \S\ref{de-projection} for clusters with
arbitrary triaxial shapes.

In \S\ref{test_cases}, we will introduce the simple analytic clusters
we use to test the de-projection method.  We will describe how we
contruct artificial observations of these clusters, and specify our
assumptions about the uncertainties in these observations in
\S\ref{artificial}.  Some additional numerical details of the
calculation are described in \S\ref{numerical}.  We apply our
de-projection method to these clusters and present our results in
\S\ref{results}.  We discuss the uncertainties in the inferred
distance, both statistical and systematic in \S\ref{discussion}. 
Finally, in \S\ref{conclusions}, we will summarize our conclusions and
highlight directions for future work.

\section{The Elongation-Distance Degeneracy}

\Label{degeneracy}

\subsection{The Spherical Case}

To understand how the distance can be determined from X-ray and SZ
effect measurements, consider the observables.  The Sunyaev-Zel'dovich
effect is due to the Compton scattering of microwave background photons
by the electrons in the intracluster plasma, with a probability per
unit path length, $dl$, equal to the Thomson cross-section, $\sigma_T$,
times the electron density, $n_e$.  The photon frequency, $\nu$,
follows a random walk with $( \delta \nu / \nu )^2$ per scattering
equal to the line of sight velocity dispersion $\langle v_{ LOS }^2
\rangle = k T_e / m_e$ of the electrons, divided by $c^2$.  The
variance $( \delta \nu / \nu )^2$ integrated along the path through the
cluster is given by the Compton $y$-parameter,
\begin{equation}
y \defeq \left ( { \delta \nu \over \nu } \right )^2 =
{ k \sigma_T \over m_e c^2 } \int d l \, n_e T_e , \label{y_parameter}
\end{equation}
where $k$ is Boltzmann's constant.  Observationally, what is measured
is not the frequency shift of individual photons, but the change in the
black body spectrum of the CMB.  This is commonly expressed in terms of
the shift in the brightness temperature at a given frequency, $( \Delta
T / T )_{ \rm CMB } ( \nu )$, which is proportional to $y$:
\begin{equation}
\left ( { \Delta T \over T } \right )_{ \rm CMB } ( \nu ) = f ( h \nu /
kT_{ \rm CMB } , T_e ) \, y , \label{define_decrement}
\end{equation}
where $f \approx - 2$ (plus relativistic corrections) when $h \nu /
kT_{ \rm CMB } < 1$.  When we refer to the SZ effect, we mean,
technically, the thermal SZ effect due to the isotropic, thermal
motions of the electrons.  If the cluster has a nonzero peculiar
velocity with respect to the Hubble flow, this will cause a systematic
frequency shift known as the kinetic SZ effect.

The X-ray emissivity is due to Bremsstrahlung from electron-ion
collisions, so it is proportional to the square of the electron
density.  Modern X-ray telescopes, such as the Chandra X-ray
Observatory (Chandra) and XMM-Newton, can measure the X-ray surface
brightness as a function of photon energy, with spectral resolving
power $E / \Delta E \sim 20$ -- $40$ at $6 \keV$. Thus, they can
measure the X-ray surface brightness per unit energy interval,
\begin{equation}
{ d \Sigma_X \over dE } = { 1 \over 4 \pi ( 1 + z )^3 } \int d l \,
n_e^2 { d \Lambda ( T_e , E^\prime ) \over dE^\prime } ,
\end{equation}
where $z$ is the redshift of the cluster, $E$ is the photon energy in
the observer's frame, $E^\prime \defeq E ( 1 + z )$ is the rest frame
photon energy, and $d \Lambda ( T_e , E^\prime ) / dE^\prime$ is the
X-ray spectral emissivity.

Both the X-ray surface brightness and the temperature decrement can be
mapped as a function of angular position, $\thetavec$, on the sky.  If
the electron density and temperature are spherically symmetric, then
these spatially resolved observations can be de-projected.  Because the
observations are functions of angular position, $\bf \theta$, the
de-projection determines $D n_e T_e$ and $D n_e^2 \, d \Lambda ( T_e ,
E^\prime ) / dE^\prime$ as functions of $r / D$, where $r$ is the
radius and $D$ is the angular diameter distance to the cluster.  Given
X-ray observations with sufficient spatial and energy resolution, the
temperature profile, $T_e ( r / D )$, can be inferred.  This allows us
to eliminate the temperature-dependent factors, to obtain $D n_e$ from
$( \Delta T / T )_{ \rm CMB }$ and $D n_e^2$ from $\Sigma_X$.  The
ratio of the square of the SZ effect measurement to the X-ray
measurement gives the angular diameter distance.

\subsection{The Elongation-Distance Degeneracy}

Now, suppose that contours of constant $n_e$ and $T_e$ are not
spherical, but are instead elongated by a factor of $Z$ along the line
of sight.  Then, the integration path length and thus the normalization
of the projected observables will be increased by a constant factor of
$Z$.  The shapes of the measured profiles, $( \Delta T / T )_{ \rm CMB
} ( { \bf \theta } )$ and $\Sigma_X ( { \bf \theta } )$ will be
unchanged, so there will be no obvious sign that the cluster is not
spherical.  If the data are analyzed under the assumption of spherical
symmetry, the inferred distance will be $D Z^2 / Z = DZ$, since the
square of the temperature decrement is being compared to the X-ray
surface brightness.

Any method which relies on the normalization of quantities integrated
along line of sight suffers from this degeneracy; only the product of
the distance and the elongation can be determined.  Observations of
additional projected quantities do not help.  The only way to break the
degeneracy is to postulate an additional constraint among the cluster
properties.  The hydrostatic equilibrium condition,
\begin{equation}
\grad p = - \rho_{ gas } \grad \phi , \label{hydrostatic}
\end{equation}
provides such a constraint between the pressure, $p$, the gas density,
$\rho_{ gas }$, and the gravitational potential, $\phi$.

Of course, while hydrostatic equilibrium adds a constraint, it also
introduces a new unknown, the gravitational potential, $\phi$, of the
cluster.  Fortunately, this extra degree of freedom can in turn be
constrained by weak gravitational lensing observations, as we will see
in \S\ref{weak_lensing}.  Then, in \S\ref{breaking}, we will show how
adding weak lensing data and the hydrostatic equilibrium condition
allows us to break the degeneracy.

\subsection{Gravitational Lensing}

\label{weak_lensing}

The gravitational potential of a cluster deflects light rays from
sources behind the cluster.  The lensing effect is determined by the
convergence, $\kappa$, which is the ratio of the projected surface
density, $\Sigma ( \thetavec )$, of the lens at angular position
$\thetavec$, to the critical surface density,

\begin{equation}
\Sigma_{ \rm cr } \defeq { c^2 \over 4 \pi G } { D_s \over D_l D_{ ls }
} , \label{critical_density}
\end{equation}
where $D_s$, $D_l$, and $D_{ ls }$ are angular diameter distances to
the source, to the lens, and from the lens to the source, respectively.
 Since each of the angular diameter distances is proportional to
$H_0^{-1}$, $\Sigma_{ \rm cr }$ is proportional to $H_0$.  However, the
product $D_l \Sigma_{ \rm cr }$ which we will encounter depends only on
the source and lens redshifts, and on the cosmological parameters
($\Omega_0$, $\Lambda_0$, etc.) which determine the shape of the
redshift-distance relation, but is independent of $H_0$.

The convergence is also equal to $( 1 / 2 ) \del^2 \psi$, where 
\begin{equation}
\psi ( \thetavec ) = { 1 \over 2 \pi G \Sigma_{ \rm cr } D_l^2 } \int d
l \, \phi , \label{lens_potential}
\end{equation}
is called the lens potential \citep{SEF92III}.

The deflection of light rays distorts the images of background
galaxies, magnifying them in both area and flux, and changing their
shapes.  In the weak lensing limit, where the second partial
derivatives $\left | \psi_{ , ij } \right | \ll 1$, these effects must
be measured statistically by averaging over many background galaxies. 
In this limit, the images are magnified by $1 + 2 \kappa$, while the
distortion in their shapes is characterized by the 2-component shear,
$\gamma_i$, where
\begin{equation}
\gamma_1 = { 1 \over 2 } ( \psi_{ , 11 } - \psi_{ , 22 } )
\end{equation}
and
\begin{equation}
\gamma_2 = \psi_{ , 12 }.
\end{equation}

Direct measurements of the magnification in the weak lensing limit are
difficult, though several techniques have been suggested \citep{BN95,
BTP95III} and at least one cluster has been detected and mapped by such
techniques \citep{Taylor98}.  More frequently, the shear is measured
from the galaxies shapes.
\Citet{KS93} demonstrated a method for reconstructing $\kappa$, up to a
constant offset, from maps of the shear \citep[see][for a
review]{Mellier99}.  The ambiguity of this offset is known as the mass
sheet degeneracy.

We should note that equation~(\ref{critical_density}) assumes a single,
known source redshift, $z_s$.   In the weak lensing limit, $\Sigma_{
\rm cr }^{-1}$ can simply be replaced with its average over the
redshifts of the sources \citep{KS93, SeitzSchneider97}.  Still, it is
necessary to estimate the source redshift distribution.  For $z_l \ll
z_s$, this task is simplified by the fact that the ratio $D_s / D_{ ls
}$ becomes independent of the source redshift.  More generally, several
methods have been suggested for estimating the source redshift
distribution from the lensing data itself \citep{Smail94, Kneib94,
Kneib96III, BN95}.

\subsection{Breaking the Degeneracy}
\label{breaking}

Equation~(\ref{hydrostatic}) implies that the contours of constant
pressure coincide with the cluster isopotentials. Taking the curl of
both sides, we see that $\grad \rho_{ gas } \cross \grad \phi = 0$, so
the isopotentials are also contours of 
constant gas density.  Finally, since $p = \rho_{ gas } kT / \mu m_p$,
where $\mu m_p$ is the mean mass per particle, they are contours of
constant temperature as well.  Thus, for a potential with a single
minimum, both $\rho_{ gas }$ and $T$ are functions of $\phi$.  Then,
the balance between pressure and gravity perpendicular to the
isopotentials can be rewritten as
\begin{equation}
{ dp \over d \phi } = - \rho_{ gas } , \Label{pressure_balance}
\end{equation}
or
\begin{equation}
{ dT \over d \phi } + T { d \log n_e \over d \phi } = - { \mu m_p \over
k }.\Label{dimensionless_pressure_balance}
\end{equation}

Since the observations are made as a function of angular position,
$\thetavec$, it is convenient to define a coordinate $\zeta = l / D$. 
Then, we can define a three-dimensional position $\xvec \defeq \rvec /
D = ( \thetavec , \zeta )$ in radians.  We also define the ellipsoidal
equivalent, $m$, of a radius, by
\begin{equation}
m^2 \defeq \xvec \cdot E^2 \cdot \xvec , \label{ellipsoidal_radius}
\end{equation}
where $E^2$ is a symmetric, positive-definite matrix.  Surfaces of
constant $m$ form a family of similar, concentric, coaxial ellipsoids. 
In the case of a cluster with isopotentials elongated along the line of
sight by $Z$, 
\begin{equation}
m = \sqrt { \thetavec^2 + \zeta^2 / Z^2 }.
\end{equation}
The gas properties are functions of $\phi$ and thus of $m$.

If we apply a spherical de-projection to such an elongated cluster, we
will obtain $DZ n_e T_e$ and $D Z n_e^2 \, d \Lambda ( T_e , E^\prime )
/ dE^\prime$ as functions of $m$.  Again, we can extract the
temperature profile, $T_e ( m )$, which depends only on the ratio
between the X-ray surface brightness in different bands, and is
therefore independent of $Z$.  Rewriting
equation~(\ref{dimensionless_pressure_balance}) in terms of derivatives
with respect to $m$, we see that
\begin{equation}
{ dT \over d m } + T { d \log n_e \over d m } = - { \mu m_p \over k } {
d \phi \over d m }.\Label{elongated_pressure}
\end{equation}
Both terms on the left-hand side of this equation can be inferred from
the X-ray observations, independent of $D$ and $Z$, so we can integrate
to find $\phi ( m )$ independent of $D$ or $Z$. 
Equation~(\ref{lens_potential}) relates $\phi$ to $\psi$, so we can
predict
\begin{equation}
\int d \zeta \, \phi \left ( m ( \thetavec , \zeta ) \right ) = { 1
\over Z } \int d ( l / D_l ) \, \phi = { 2 \pi G \Sigma_{ \rm cr } D_l
\over Z } \psi ( \thetavec ) \label{predicted_lensing}
\end{equation}
as a function of $\thetavec$.  As we noted in \S\ref{weak_lensing},
$\Sigma_{ \rm cr } D_l$ depends upon the source and lens redshifts and
on cosmology, but is independent of $D_l$ and $H_0$, so we can predict
$\psi / Z$.
Taking the appropriate partial derivatives with respect to $\theta_i$,
we can calculate both $\kappa / Z$ and $\gamma / Z$.  Comparing with
$\kappa$ or $\gamma$ from lensing observations, we can solve for the
elongation, $Z$.  Either $\kappa$ or $\gamma$ is sufficient to fix $Z$.
 Therefore, in principle, $Z$ can be determined from weak shear
measurements alone, without the need to infer $\kappa$ or to resolve
the mass sheet degeneracy.

Thus, given the X-ray and weak lensing observations, the hydrostatic
equilibrium condition, the cosmology and the distribution of source
redshifts, we can infer the elongation, $Z$.  This breaks the
degeneracy. Recalling that the X-ray and SZ effect observations
determine $DZ$, we see that we can infer $D$ and thus $H_0$.

\section{Cluster De-projection}

\label{de-projection}

Finding the three-dimensional cluster properties from the
two-dimensional observations is an inverse problem.  Following the
usual approach, we construct a set of parameterized three-dimensional
cluster models.  The angular diameter distance to the cluster is
included as one of the parameters.  For any given set of parameter
values, we can predict the observations from the model.  We define a
$\chi^2$ function quantifying the differences between the actual
observations and these predictions, and minimize it to find the best
fit parameters.

As always, the trick is to choose the set of models well.  If the set
is not general enough, we will get deceptively tight constraints on the
distance, based as much on the restricted choice of models as on the
data.  Any systematic differences between the best-fitting model and
the data may also translate into systematic errors in our estimation of
the distance and elongation.  On the other hand, if the set is too
general, it will be impossible to constrain the parameters.

To begin, we will discuss the appropriate symmetry assumptions for
elongated clusters in \S\ref{symmetry}.  We will argue that an
ellipsoidal mass distribution is the best choice.  In
\S\ref{mass_model}, we will discuss the constraints on the mass
distribution from the weak gravitational lensing observations.  We will
then add a hydrostatic equilibrium model for the gas and define the
$\chi^2$ function in \S\ref{gas_model}

\subsection{De-projection and Symmetry}

\Label{symmetry}

Projection always involves loss of information.  Therefore,
de-projection requires symmetry assumptions.  Since we are interested
in elongated clusters, it is natural to consider ellipsoidal symmetry.
Analytically, the de-projection is simplest if the gas properties, and
thus the gravitational potential, are constant on a family of similar,
concentric, coaxial ellipsoids.  Unfortunately, there is no guarantee
that a given ellipsoidal potential will correspond to a physically
sensible mass density distribution, $\rho = \del^2 \phi / 4 \pi G$.  In
particular, for isothermal gas, the ellipsoidal generalization of the
usual $\beta$-model for cluster X-ray emissivity has a potential
\begin{equation}
\phi \propto kT \log ( 1 + m^2 ).
\end{equation}
However, for prolate ellipsoids with axis ratio greater than $\sqrt { 3
/ 2 }$, the corresponding density distribution becomes dumbbell-shaped,
which is unlikely for a cluster in equilibrium.  For oblate ellipsoids
with axis ratio less than $\sqrt { 1 / 2 }$, the density actually
becomes negative in some regions.  Thus, while the ellipsoidal
potential model was useful for demonstrating the distance-elongation
degeneracy and its resolution analytically in \S\ref{degeneracy}, it is
not necessarily the best model to use in practice.

Therefore, we will instead assume that the total mass density is
constant on similar ellipsoids.  This ensures a sensible mass
distribution.  However, it also means that the isopotentials will not
in general be ellipsoids.  In addition, the shapes of the isopotentials
will depend not only on the shape of the isodensity contours, but on
the density profile, $\rho ( m^2 )$, as well.  Thus, we cannot
construct a model to predict the X-ray and SZ effect observations until
we have determined the density profile.  Consequently, we must break
the construction of the model into two steps.  First, we construct a
set of possible mass models, including both the shape and the density
profile, as constrained by the gravitational lensing observations.  For
each mass model, we calculate the gravitational potential. Second, we
add the intracluster gas in hydrostatic equilibrium.  This allows us to
predict the X-ray and SZ effect observations, and to adjust the free
parameters of both the mass and gas models to find the best fit.
  
\Citet{JS2002} have fit triaxial models to the mass within isodensity
contours of simulated clusters.  They find that the orientation of the
axes is relatively stable as a function of density, though the central
regions tend to be more elongated.  However, to keep the number of
model parameters to a minimum, we use ellipsoids of fixed shape and
orientation.

\subsection{Gravitational Lensing and the Mass Model}

\label{mass_model}

The shape of a triaxial ellipsoidal mass distribution is described by a
pair of axis ratios. The orientation relative to the observer's line of
sight is specified by three Euler angles.  These five parameters,
together with the density profile, $\rho ( m^2 )$, characterize the
mass distribution.  There is an additional arbitrary choice of scale in
the definition of $m^2$.  However, a re-scaling of $\rho ( m^2 )$ can
always compensate for a change in the choice of scale, so the scale is
not an additional parameter. 

The weak gravitational lensing observations measure the convergence,
$\kappa ( \thetavec ) = \Sigma ( \thetavec ) / \Sigma_{ \rm cr }$.  The
critical surface density depends on $H_0$, but the product $D_l
\Sigma_{ \rm cr }$ does not. Thus, we can measure the product 
\begin{equation}
D_l \Sigma ( \thetavec ) = D_l \int d l \, \rho.
\end{equation}

For an ellipsoidal mass distribution, $\rho ( m^2 )$, the projected
mass density will be constant on a family of similar ellipses,
parameterized by a single axis ratio and a single orientation angle. 
By fitting an elliptical projected mass distribution to the lensing
data, we obtain these two parameters, together with $D_l \Sigma$ as a
function of the elliptical coordinate, $b$, defined analogously to the
ellipsoidal ``radius'', $m$.  Uncertainties in the lensing data will of
course allow for some variation in these quantities.  For now, however,
we will assume that they are fixed.

Since the shape and orientation of the ellipsoidal distribution are
described by five parameters, there will still be three free parameters
required to uniquely determine the shape and orientation of the mass
distribution, even with perfect lensing information.  First, consider a
particular member of this three-dimensional parameter space; suppose
that the surfaces of constant density are prolate axisymmetric
ellipsoids generated by rotating the elliptical contours of projected
mass about their major axis. Then, we can de-project to find the
density profile $D_l^2 \rho ( m^2 )$ in the same way one de-projects a
spherical distribution.

To obtain the rest of the possible ellipsoidal de-projections, we
consider the three families of transformations which map ellipsoids
into ellipsoids without changing the projected mass distribution. 
Specifically, one family corresponds to a stretching along the line of
sight,
\begin{equation}
\rho ( \xvec ) \rightarrow \rho^\prime ( \xvec ) =
Z_{ \rm mass }^{-1} \rho ( \xvec^\prime ) ,
\end{equation}
where
\begin{equation}
\xvec^\prime = ( \thetavec^\prime , \zeta^\prime ) = ( \thetavec ,
\zeta / Z_{ \rm mass } ) ,
\end{equation}
and $Z_{ \rm mass }$ is the elongation of the isodensity contours
relative to the prolate case.

The other two families correspond to shear transformations,
\begin{equation}
\rho ( \xvec ) \rightarrow \rho^\prime ( \xvec ) = \rho ( \xvec^\prime
) ,
\end{equation}
where
\begin{equation}
\xvec^\prime = ( \thetavec^\prime , \zeta^\prime ) = ( \thetavec ,
\zeta + \svec \cdot \thetavec ) ,
\end{equation}
characterized by a $2$-vector, $\svec$, in which the mass is displaced
along the line of sight by an amount, $\svec \cdot \thetavec$,
proportional to $\thetavec$.

The stretching transformation does not commute with the shearing
transformation, so we must specify the order in which they are applied.
 For a given elliptical projected mass distribution, we obtain a
particular de-projection from the canonical prolate case by shearing it
by $\svec$, and then stretching it along the line of sight by $Z_{ \rm
mass }$.

We will test various values of $Z_{ \rm mass }$ and $\svec$, and use
the X-ray and SZ effect observations to determine the best values.  All
the projected observables are invariant under a reflection through the
plane of the sky, which corresponds to $\svec \rightarrow - \svec$, so
it is impossible to tell which end of the cluster is pointing at us.  

Ellipsoidal clusters axisymmetric about the line of sight, for which
$\svec = 0$, have an axis ratio equal to $Z_{ \rm mass }$.  In the
general case of triaxial ellipsoids of arbitrary orientation, that will
no longer be true.  The axis ratios will depend in a complicated manner
on the axis ratio of the elliptical projected mass distribution, the
magnitude and orientation (with respect to the projected ellipse) of
the shear, $\svec$, and on $Z_{ \rm mass }$.

To be more precise, the shear transformation preserves the thickness,
\begin{equation}
\Delta \zeta ( m , \thetavec ) = 2 \sqrt { m^2 - \thetavec \cdot S
\cdot \thetavec } ,
\end{equation}
of the canonical prolate ellipsoid at fixed $\thetavec$.  However, it
increases the total extent,
\begin{equation}
\Delta \zeta ( m ) \defeq \max_{ \thetavec } \zeta ( m , \thetavec ) -
\min_{ \thetavec } \zeta ( m , \thetavec ) = 2 \max_{ \thetavec } \zeta
( m , \thetavec ) ,
\end{equation}
in $\zeta$ spanned by the ellipsoid.  Thus an ellipsoid with $Z_{ \rm
mass } = 1$ has the same thickness as the canonical prolate ellipsoid,
but can have a much larger extent in the $\zeta$ direction.  The
subsequent stretching increases both the thickness, $\Delta \zeta ( m ,
\thetavec )$, and the extent, $\Delta \zeta ( m )$, by a factor of $Z_{
\rm mass }$.  Nonetheless, for lack of a better term, we will continue
to refer to $Z_{ \rm mass }$ as the line of sight elongation.

Finally, note that the isopotentials of an ellipsoidal mass
distribution are rounder than the isodensity contours (and are in
general not ellipsoids).  Thus, even in the $\svec = 0$ case, the
typical elongation, $Z$ of the isopotentials, which determines the
ratio of true distance to that inferred from the spherical method, will
be smaller than the axis ratio of the mass distribution.  Conversely,
for clusters in hydrostatic equilibrium, the axis ratio of the mass
distribution will be greater than the typical values estimated from the
shapes of X-ray isophotes.

The calculation of the potential of an ellipsoidal mass distribution
has been extensively discussed \citep[see, e.g.,][]{ChEFE, BT87}. 
Since we are working in coordinates in radians,
\begin{equation}
\grad_{ \xvec }^2 \phi ( \xvec ) = D_l^2 \grad_{ \rvec }^2 \Phi ( \rvec
) = 4 \pi G \rho ( \xvec ) D_l^2
\end{equation}
so the potential, $\phi ( \xvec )$, is defined in terms of the
combination $\rho ( \xvec ) D_l^2$ which is determined by the
observations, independent of $H_0$.  It is convenient to choose the
zero point of the potential at the center of the cluster.  Note that
the hydrostatic equilibrium condition, equation~(\ref{hydrostatic}),
has the same form in terms of the dimensionless $\xvec$, as long as the
pressure is also differentiated with respect to $\xvec$.

\subsection{Gas Model and the Chi-Squared Function}

\label{gas_model}

Once we have the cluster potential, $\phi ( \xvec )$, we can add the
intracluster gas.  In hydrostatic equilibrium, the gas can be described
by its temperature profile, $T ( \phi )$ and the central electron
density, $n_{ e 0 }$.  To minimize $\chi^2$ numerically, we need a
discrete parameterization for the temperature profile.  We let $\log T
( \phi )$ be a polynomial of fixed order in $\phi$.

We then integrate equation~(\ref{dimensionless_pressure_balance}) to
obtain the density profile, $n_e ( \phi )$.  From the density and
temperature profiles, we can calculate the electron pressure and X-ray
emissivity in each energy band.  Using $\phi ( \xvec )$, and including
a factor of $D$ for the path length, we can project to predict the SZ
effect temperature decrement and the X-ray surface brightness.  Given a
model of the instruments used to observe the cluster, we can convert
these predictions to the measured quantities.

The predictions depend on the various parameters of the model.  They
depend nonlinearly on the elongation and shear, $Z_{ \rm mass }$ and
$\svec$, which fix the shape and normalization of the deprojected mass
distribution, as well as on the central electron density, $n_{ e 0 }$,
and the temperature profile, $T ( \phi )$.  Finally, they are
proportional to the angular diameter distance, $D$, to the cluster.

We construct a $\chi^2$ function for the X-ray and SZ effect
measurements in the usual way, weighting the square of the differences
between the observations and the model predictions by the reciprocal of
the variance of each measurement.  We apply the chain rule to calculate
the partial derivatives of $\chi^2$ with respect to the gas parameters
and the distance.  The dependence on the shape of the cluster is more
complicated.  We therefore break the $\chi^2$ minimization problem into
an inner part and an outer part.  For a fixed set of shape parameters,
we minimize with respect to the gas parameters and distance to find the
best fit for that shape.  This minimization defines an integrated
$\chi^2$, 
\begin{equation}
\chi^2 ( Z_{ \rm mass } , \svec ) \defeq \min_{ n_{ e 0 } , T ( \phi )
, D } \chi^2 \left ( Z_{ \rm mass } , \svec , n_{ e 0 } , T ( \phi ) ,
D \right ).
\end{equation}
Then, we minimize this integrated $\chi^2$ with respect to the
elongation and shear, $Z_{ \rm mass }$ and $\svec$, which determine the
shape of the cluster.  For the inner minimization with respect to the
gas parameters and distance, we use the Levenberg-Marquardt nonlinear
chi-squared minimization algorithm, which takes advantage of the
derivative information.  

Note that the lensing observations are not included in the $\chi^2$
function. We already had to use them to fix the projected mass density,
and thus determine $\phi$, up to the three remaining shape parameters. 
Without this, we could not have made predictions for the projected gas
observables.  This special role for the lensing data is a direct
consequence of our choice of symmetry assumption for the mass
distribution; the shape of the isopotentials contours depends on the
mass profile, and not just on the shape of the mass distribution.

As a result, the effect of uncertainties in the lensing observations
must be considered separately.  They will lead to some range of
acceptable parameters in the fit to the projected mass information in
\S\ref{mass_model}.  By repeating the inversion procedure for different
values of these parameters, we can test the effect on the angular
diameter distance.

\section{Test Cases}
\label{test_cases}

To test our de-projection method, we apply it to artificial
observations of model clusters, to see how well we recover the original
shape and distance.  We will refer to these clusters as the true
clusters.  Our true clusters, like the models used to fit them, have
ellipsoidal mass distributions, so that we know a priori what the best
fit shape ought to be in the absence of noise.  The purpose of this
test is to determine how sensitive the method is to the uncertainties
in the observations.  In particular, we want to ensure that there were
no additional degeneracies which our analytic treatment in
\S\ref{degeneracy} might have missed.

For the mass density profile, $\rho$, as a function of our angular
ellipsoidal coordinate, $m$, we use a $\beta$-model,
\begin{equation}
\rho ( m ) = \rho ( 0 ) \left [ 1 + { m^2 \over m_0^2 } \right ]^{ - 3
\beta / 2 }
\end{equation}
with core $m_0$.  We choose $\beta = 2 / 3$.  While such a model with a
core is generally more appropriate to the X-ray gas than to the cluster
matter distribution, it should be sufficient for testing purposes. 
This choice also reduces the number of integrals which must be done
numerically to calculate the potential.

For the X-ray gas, we assume a polytropic temperature profile with a
central temperature $T = 8.0 \keV$ and a polytropic index, $n$, equal
to $1.1$.  We assume a fully ionized hydrogen-helium plasma with a
hydrogen fraction of $0.76$ by mass.  We take the central electron
density, $n_{ e 0 }$, equal to $1.0 \times 10^{ - 2 } \cm^{ - 3 }$. 
The density normalization only affects the signal-to-noise of the gas
observations.

We use two sets of true clusters.  First, we consider ellipsoidal
clusters axisymmetric about the line of sight, which we fit with models
with the same symmetry.  This allows us to treat the observables as
functions of one variable, the angular radius, $\theta$.  It also
reduces the shape parameter space from three dimensions to one, the
mass elongation, $Z_{ \rm mass }$ relative to spherical.  Both changes
greatly reduce the CPU time required for the inversion.  We consider
clusters with different elongations, $Z_{ \rm mass }$, in the mass
distribution, relative to spherical, ranging from $0.6$ to $1.4$.  We
normalize each density profile to produce the same projected mass,
regardless of the elongation.  Thus, we are testing whether the X-ray
and SZ effect measurements allow us to distinguish between clusters
with identical lensing observations.  Recall that the elongation, $Z$,
of the potential, which directly affects the inferred distance, is
generally smaller than $Z_{ \rm mass }$.

Second, we consider the more general case of clusters of arbitrary
shape and orientation.  For this case, we use three true clusters: two
triaxial clusters and one axisymmetric but with its axis inclined to
the line of sight.  Their shapes are listed in
Table~\ref{test_ellipsoids}.  We observe the two triaxial clusters
along two different lines of sight, and the axisymmetric cluster along
one.  Table~\ref{trial_observations} lists the trial observations.  The
lines of sight are specified by spherical coordinates: the inclination,
$i$, with respect to the third ($c$) axis, and the azimuthal angle,
$\phi$, with respect to the first ($a$) axis.

\begin{table}[tb]
\caption {Triaxial Test Cases\label{test_ellipsoids}}
\begin{tabular} {llll}
\noalign{\vskip4pt}
Shape name & \multispan3 {\hfill Axes \hfill} \\
& \multispan3 {\hfill ($h^{-1} \Mpc$)\hfill} \\
& $a$ & $b$ & $c$ \\
\tableline
A & $1.2$ & $0.8$ & $2.0$ \\
B & $2.0$ & $1.0$ & $1.0$ \\
C & $0.7$ & $1.4$ & $2.0$ \\
\end{tabular}
\end{table}

\begin{table}[tb]
\caption {Trial Observations of Triaxial Test
Cases\label{trial_observations}}
\begin{tabular} {llll}
\noalign{\vskip4pt}
Observation & Shape\tablenotemark{a} & \multispan2 {\hfill Line of
sight direction \hfill} \\
& & inclination $i$\tablenotemark{b} & azimuthal angle
$\phi$\tablenotemark{c} \\
 &  & (deg) & (deg) \\
\tableline
A1 & A & $30$ & $40$ \\
A2 & A & $60$ & $75$ \\
B   & B & $60$ & $0$ \\
C1 & C & $30$ & $40$ \\
C2 & C & $60$ & $75$ \\
\end{tabular}
\tablenotetext{a}{see Table~\ref{test_ellipsoids}}
\tablenotetext{b}{with respect to the $c$ axis of the ellipsoid}
\tablenotetext{c}{with respect to the $a$ axis of the ellipsoid}
\end{table}

When choosing the model length and density scales, it is sometimes
easier to think in terms of the physical quantities.  We choose the
physical core length scale $m_0 D$ to be $0.1 \, h^{-1} \Mpc$.  For the
clusters axisymmetric about the line of sight, we choose a central
density, $\rho ( 0 )$, equal to $4 \times 10^{ 15 } Z_{ \rm mass }^{-1}
\, h^2 \MSun / \Mpc^3$.  For the other clusters, $\rho ( 0 ) = 4 \times
10^{ 15 } \, h^2 \MSun / \Mpc^3$, regardless of shape and orientation.

Of course, to convert these physical quantities to the
$H_0$-independent quantities actually used in the model, we need to
specify the scaled distance, $h D$.  We will take the redshift, $z$, of
our clusters to be $0.2$.  This is far enough away to make weak lensing
measurements feasible, yet close enough for X-ray instruments such as
Chandra to produce high signal-to-noise maps in many energy bands.  We
assume a flat $\Lambda$CDM cosmology with $\Omega_0 = 0.3$ and
$\Lambda_0 = 0.7$, which gives $D = 476 \, h^{-1} \Mpc$.  Then, we
calculate $m_0$ and $\rho ( 0 ) D^2$.

Note that for a given choice of the physical quantities, the angular
scale of the cluster, will vary slightly with cosmology, as will the
normalizations of the X-ray and SZ effect observations.  Thus, the
signal-to-noise of these observations does depend slightly on our
choice of cosmology.  However, the angular scale and signal-to-noise
would also vary from cluster to cluster in a given cosmology.  In a
different cosmology, the same calculation with the same angular scale
and signal-to-noise would simply correspond to a slightly different
physical cluster, but would not be fundamentally altered.

When we project the local electron pressure and X-ray emissivity to
create the artificial observations of the intracluster gas, we also
need the actual distance, $D$.  We take $h = 0.7$, so $D = 681 \Mpc$. 
Again, given the choice of the physical mass profile parameters, this
will affect the signal-to-noise.  Of course, neither this assumed value
of $h$ nor that of $D$ is used by the de-projection procedure.

\section{Artificial Observations}
\label{artificial}

Next, we need to create artificial observations of these cluster
models, both the true clusters and the trial models used in the
$\chi^2$ fitting.  Our goal is to see how the observational noise and
uncertainties affect our ability to recover the true shape and
distance.  We do not attempt to model in detail all of the properties
of specific instruments.  Instead, we use simplified models with
roughly the same sensitivity.  The possibility of creating high
signal-to-noise weak lensing and SZ effect maps is relatively new. 
Therefore, we allow for some improvement over the current
state-of-the-art in these observations. The field of X-ray astronomy is
comparatively mature, so we base our model more closely on current
instrumentation.  We add noise to the true cluster observations but not
to the trial models.  Otherwise, we use the same models to create both
sets of artificial observations.

\subsection{Weak Lensing}

\label{lensing_uncertainties}

As we mentioned in \S\ref{gas_model}, incorporating uncertainties in
the weak gravitational lensing measurements requires additional Monte
Carlo testing of the entire de-projection analysis.  Because of this,
as well as rapid advances in the weak lensing observations, we have
entirely ignored the weak lensing uncertainties in our initial tests. 
Instead, we use the exact de-projection of the density profile of the
true cluster.  In doing so, we are also not restricting ourselves to
using only the weak shear information.  Thus, while we showed in
\S\ref{breaking} that the elongation could be recovered from the weak
shear alone, our numerical tests do not demonstrate this.

\subsection{SZ Effect}

We base our artificial SZ effect observations roughly on the maps made
by fitting the Owens Valley Radio Observatory (OVRO) millimeter-wave
interferometer with $30 \GHz$ ($1 \cm$ wavelength) receivers
\citep{CJG96}.  At this frequency, the proportionality constant between
the temperature decrement and the Compton $y$-parameter in
equation~(\ref{define_decrement}) differs slightly from $- 2$. 
Relativistic corrections \citep{Rephaeli95, CL98} also modify the
proportionality constant.  However, the total correction is roughly $5
\%$ \citep{Patel2000}, so we ignore it.

A two-element interferometer measures the complex visibility,
\begin{equation}
V ( \uvec ) = \int d \thetavec \, I ( \thetavec ) \exp \left ( - 2 \pi
i \uvec \cdot \thetavec \right ) , \label{visibility}
\end{equation}
a Fourier component of the intensity, $I$, where $\uvec = \Delta \rvec
/ \lambda$ is the baseline separation of the antenna elements in units
of the wavelength, $\lambda$.    Equation~(\ref{visibility}) is valid
as long as the separation, $\Delta \rvec$, between the antennas is much
larger than their diameter, which is usually the case in radio
interferometry.  The Fourier transform of this sparse Fourier-space
sampling is called the ``dirty'' beam.  Taking the Fourier transform of
the visibilities results in a ``dirty'' image which is the convolution
of the true intensity with the dirty beam.  Convolution techniques are
used to recover a smooth, ``clean'' image with the equivalent
resolution.  Clusters of galaxies, however, are large compared to the
angular resolution of interferometric arrays. The ability to measure
the temperature decrement on large scales is limited by the shortest
baselines.  Therefore, these observations are generally made with the
array in a nearly close-packed configuration to maximize sensitivity to
the SZ temperature decrement on cluster scales, with some longer
baselines for detection and removal of point sources.  For OVRO, this
system produces a roughly $1 \arcmin$ beam and maps with an rms noise
of $\sim 27 \muK$ per beam in the temperature decrement with $60
\hours$ of observations \citep{Patel2000}.  We adopt this resolution
and sensitivity.  

Such close-packed configurations produce ``dirty'' beams which are
actually smooth, so the data can be analyzed in real space
\citep{PNKS2002}.   The temperature decrement measurements have
Gaussian white noise, so the noise remains uniform and Gaussian when
transformed to real space.  The physical separation of the dishes must
be greater than the diameter of the individual dishes.  This limits the
shortest baselines measured, which results in a hole in the Fourier
space coverage.  In real space, the effect of this hole is
approximately equivalent to subtracting the diffraction limited image
which would be produced by a single dish from an image with the
diffraction limit of the entire array.  This gives an effective beam
which has a total area of zero, as appropriate for interferometric
observations with no measurements at zero-spacing.  For the OVRO array,
each dish has a $10.4 \meters$ diameter, corresponding to a diffraction
limit of $4 \arcmin$ at $30 \GHz$.

For simplicity, in the axisymmetric case, we limit the field of view of
our temperature decrement observations to $3 \arcmin$ in radius,
covered with three $1 \arcmin$ annuli, and remove the short baselines
by subtracting the average temperature decrement over this field from
the individual annuli.  This simple approach produces the correct
spatial resolution and noise per pixel, but it reduces the field of
view and the total number of pixels.   It also ties the diameter of the
field of view to the dish size, preventing us from changing one without
changing the other.

For clusters with arbitrary orientation, the field of view is no longer
circular, so this approach no longer makes sense.  Therefore, we use an
$8 \arcmin \times 8 \arcmin$ field of view, with a rectangular grid of
$1 \arcmin$ pixels, to represent the image with the diffraction limit
of the entire array.  We then convolve this image with a Gaussian
of unit area and full width at half maximum equal to the single dish
diffraction limit and subtract the convolved image from the original.

\subsection{X-ray Observations}

For our artificial X-ray observations, we use a simplified model of the
Advanced CCD Imaging Spectrometer (ACIS) on the Chandra X-ray
Observatory as described by the \citet{CPOG}.  The four
front-illuminated CCD chips of the ACIS-I imaging configuration have
$0.492 \arcsec$ pixels over a $16.9 \arcmin \times 16.9 \arcmin$ field
of view.  The half-energy radius varies roughly from 1 to 10 pixels
from the center of the field of view to an off-axis angle of $8
\arcmin$.  The full spatial resolution is unnecessary for our purposes,
so in the axisymmetric case, we use $50$ annuli, each $9.6 \arcsec$
wide, out to a radius of $8 \arcmin$.  In the case of the clusters of
arbitrary orientation, we use a $20^2$ rectangular grid covering and $8
\arcmin \times 8 \arcmin$ field of view (one-quarter of the Chandra
ACIS-I field), giving $24 \arcsec \times 24 \arcsec$ pixels.  This
resolution is quite crude, but is necessary to ensure that the
chi-squared fitting can be performed in a reasonable amount of CPU
time.

For simplicity, we ignore X-ray line emission and consider only thermal
Bremsstrahlung.  The ACIS energy resolution varies with energy and
position on the chips.  However it is generally $\lesssim 0.3 \keV$. 
We create maps in $30$ energy bands of this width between $1 \keV$ and
$10 \keV$.  To calculate the count rates, we multiply the X-ray surface
brightness for each band by the angular area of each annulus and by the
effective area of each band.  We assume an effective area of $A = 500
\cm^2$ for $E < 2 \keV$, $300 \cm^2$ for $2 < E < 5.5 \keV$, and a
linear falloff in $\log A$ to $14.3 \cm^2$ from $5.5 \keV$ to $10
\keV$.  This approximates the on-axis effective area of the
front-illuminated CCDs, while simplifying the detailed edge structure. 
For simplicity, we also ignore the change in effective area with
off-axis angle, which is roughly a $15 \%$ decrease at the edge of our
field of view.

We add a contribution to the count rates from the background (both the
actual diffuse X-ray background and the charged particle background) of
$0.07 \, { \rm counts } / { \rm chip } / { \rm \seconds }$ for $E < 2
\keV$ and $0.15 \, { \rm counts } / { \rm chip } / { \rm \seconds }$
for energies from $2 \keV$ to $10 \keV$ as estimated from the on-orbit
measured background for the ACIS-I chips \citep{CPOG}.  For simplicity,
we add the same background to the count rates for the trial models,
rather than adding a background term whose normalization is a free
parameter of the fit.

We multiply the count rates by an assumed $100 \ksec$ observation time
to find the expected number of counts in each band and annulus, and add
Poisson fluctuations to these counts.  Note that when we compute the
Poisson uncertainties in the number of counts for the $\chi^2$
function, we use the mean values.  With actual observations, only the
measured number of counts including noise would be known.  This would
introduce a bias to actual observations, which would have to be
corrected.

\section{Numerical Details}
\label{numerical}

The projected gas observables, the X-ray surface brightness and the
temperature decrement, are approximated by discrete sums as described
in the Appendix.

Our true clusters have ellipsoidal mass distributions, so the projected
mass distributions are elliptical.  Since we are not including any
uncertainties in the lensing information, it is not necessary to fit an
elliptical model to the projected mass and de-project it explicitly for
each assumed shape.  The de-projected mass profile will differ from the
true profile, $\rho ( m^2 )$, only by the ratio of the true mass
elongation, $Z_{ \rm mass }$, to the assumed value.

For clusters axisymmetric about the line of sight, our shape parameter
space is reduced to a single dimension, the elongation, $Z_{ \rm mass
}$.  For simplicity, we perform the outer minimization of $\chi^2$ with
respect to elongation by a simple grid search.  We evaluate the
integrated $\chi^2 ( Z_{ \rm mass } )$ on a uniform grid of 33 values
of assumed elongation in a range of $\pm 0.2$ about the true
elongation, and pick the elongation which gives the lowest chi-squared.

In the case of clusters of arbitrary shape and orientation, such a
brute force grid search is no longer feasible.  Instead, we use a
general-purpose multidimensional minimization routine based on Powell's
method.  To ensure that the minimization routine has converged, we
restart it twice at the previous minimum.  In practice, there was no
significant change in the solution after the first restart.

To find the initial guess for this minimization routine, we perform a
low resolution grid search over a 4 by 2 by 2 grid of cells.  We
initially evaluate the integrated $\chi^2$ at the center of each cell
and find the cell with the smallest value.  We then refine the grid by
evaluating $\chi^2$ in a 3 by 3 by 3 grid within this cell.  The shape
with the smallest integrated $\chi^2$ is used as the initial guess for
the minimization routine.

As mentioned earlier, we need a discrete parameterization for the
temperature profile.  We take $\log T ( \phi )$ be a polynomial of
fixed order in $\phi$. in the axisymmetric case, we used a fourth order
polynomial.  In the general case, we found that a third order
polynomial was more effective in avoiding unphysical oscillations in
the temperature profile in the initial guess for the cluster shape was
poor.

The inner minimization, with respect to the gas parameters and the
distance, also requires an initial guess.  We 
guess a constant temperature profile of $7.7 \keV$, roughly equal to
the true central temperature of $8 \keV$.  In the axisymmetric case, we
take the central electron density to be twice the true value, and the
distance to be $0.833 Z_{ \rm mass }^{-1} ( { \rm assumed } )$ of the
true value.  In the general case, we take the central density to be
$1.5$ times the true value, and the distance to be $1.1$ times the true
value.  As long as the initial guess for the temperature is not too
extreme, the code rapidly converges to essentially the same best fit
set of parameters, regardless of the choice of initial guess.  Using a
fairly accurate initial guess for the central temperature is realistic
since the central temperature of a relaxed cluster should be reasonably
well determined from the overall emission-weighted spectrum.

For each test case, we create 20 realizations of the artificial
observations with noise, which are analyzed independently.  This allows
us to estimate the uncertainties in the parameters inferred from a
single observation due to statistical uncertainties in the
measurements.

\section{Results}
\label{results}

\subsection{Axisymmetric Clusters}

First, we show the results of fitting the true clusters which are
axisymmetric about the line of sight, with models of the same symmetry.
 Figure~\ref{inferred_elongation} shows the mean values of the inferred
mass elongation for each true elongation.  The error bars are the $1
\sigma$ uncertainties for a single realization, calculated from the
sample variance of the 20 realizations.  The diagonal line shows $Z_{
\rm mass } ( { \rm inferred } ) = Z_{ \rm mass } ( { \rm true } )$. 
The mean elongations are consistent with no bias relative to the true
elongations (with a $3 \sigma$ upper limit of $1$ to $2 \%$).  The
uncertainties for a single realization range from $1$ to $3 \%$.  Since
we plot mean values (to test for systematic bias), the deviations from
the true elongation are of course much smaller than the error bars,
which represent the uncertainty in a single realization.

\begin{figure}
\plotone{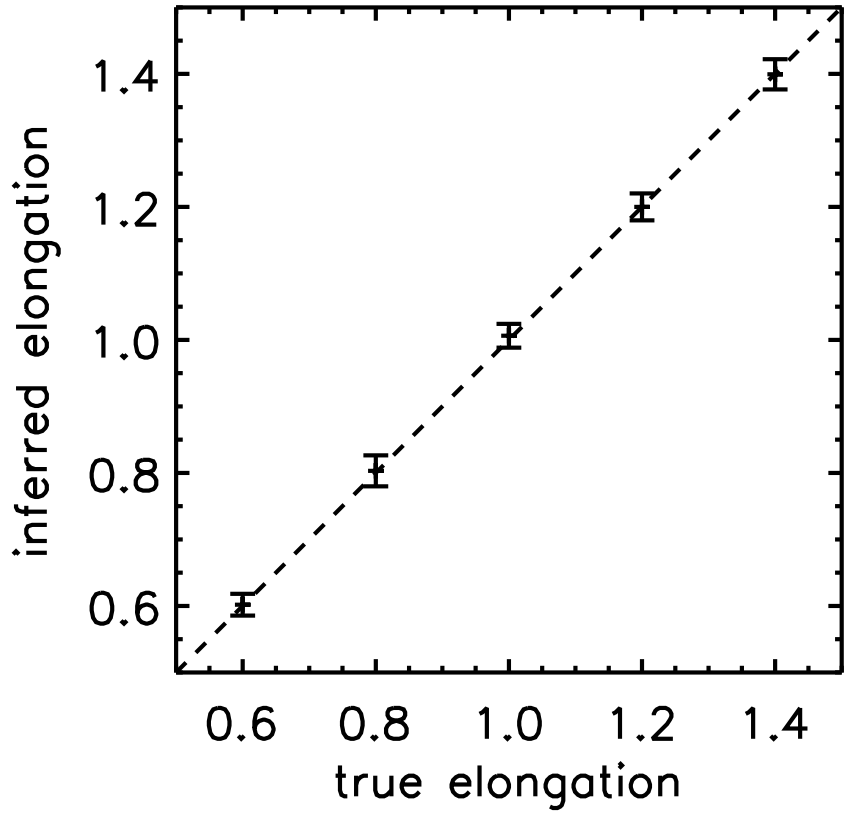}
\caption {Inferred vs. true mass elongation, $Z_{ \rm mass }$, as
estimated from 20 realizations of the artificial observations.  The
error bars are the sample standard deviation of the 20 realizations,
corresponding to the $1 \sigma$ uncertainties for a single realization.
 \Label{inferred_elongation}}
\end{figure}

The inferred elongations shown are from simultaneous fits for the
elongation, distance, central density, and temperature profile. 
However, we obtain nearly identical results if we use only the X-ray
and lensing data, and fit for elongation, temperature profile, and the
combination $n_{ e 0 }^2 D$.  This is to be expected from our analytic
analysis in \S\ref{breaking}, where we saw that constraining elongation
required the absolute temperature and the relative temperature and
density profiles.  The SZ effect by itself only constrains the product
of the two profiles.  Nor can it constrain the absolute temperature, if
$H_0$ and thus the cluster distance are unknown.  Therefore, the
elongation is likely to remain determined primarily by the X-ray and
lensing observations, even with improvements in the temperature
decrement measurements.

In Figure~\ref{inferred_distance}, we show the inferred distance for
clusters of each true elongation.  Again, the values and error bars
represent the mean over the 20 realizations, and the $1 \sigma$
uncertainties of a single realization.  The dashed line indicates the
true distance of $681 \Mpc$.  There is no significant bias (with a $3
\sigma$ upper limit of $1$ to $2 \%$), and the uncertainties in a
single realization are of order $2$ to $3 \%$.

\begin{figure}
\plotone{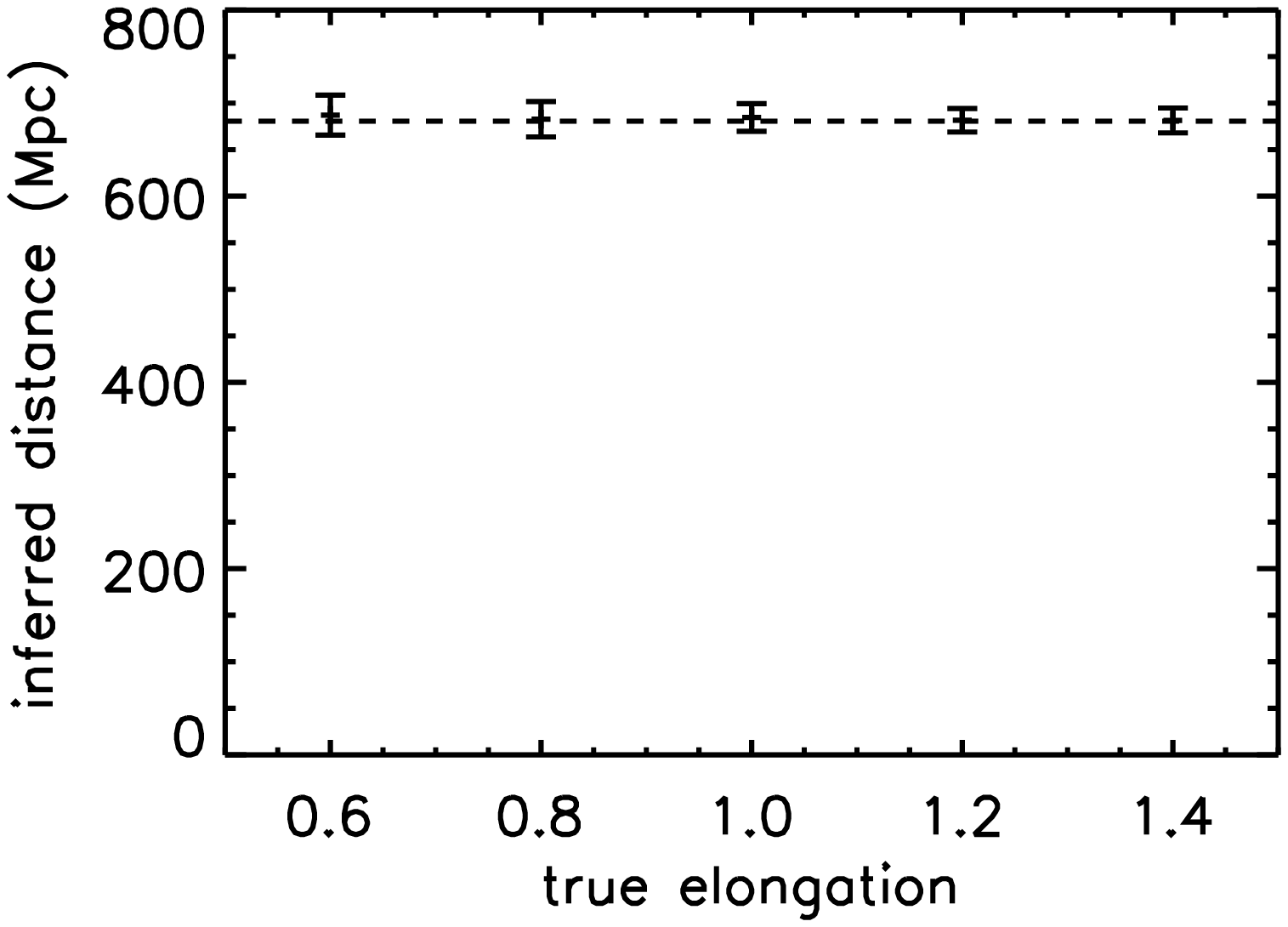}
\caption {Inferred distance for each true mass elongation, $Z_{ \rm
mass }$ as estimated from 20 realizations of the artificial
observations.  The error bars are the sample standard deviation of the
20 realizations, corresponding to the $1 \sigma$ uncertainties for a
single realization.  \Label{inferred_distance}}
\end{figure}

\subsection{General Case}

Next, we fit triaxial models to the second set of true clusters
described in \S\ref{test_cases}, those with arbitrary orientation and
shape.  For each observation listed in Table~\ref{trial_observations},
we created 20 realizations of the noise, and analyzed each one
independently.  Figure~\ref{model_distances} shows the inferred angular
diameter distance for each observation.  The values represent the mean
of the 20 realizations.  The error bars are the $1 \sigma$
uncertainties due to instrumental noise for a single realization,
calculated from the sample variance, which range from $3$ to $6 \%$. 
The dashed line indicates the true distance of $681 \Mpc$.  The mean
distances are consistent with no bias, with $3 \sigma$ upper limits
ranging from $2$ to $4 \%$.

The true elongation, $Z_{ \rm mass }$ is recovered to $2$ to $4 \%$. 
The absolute errors in the shear ranged from $0.05$ to $0.10$.

\begin{figure}
\plotone{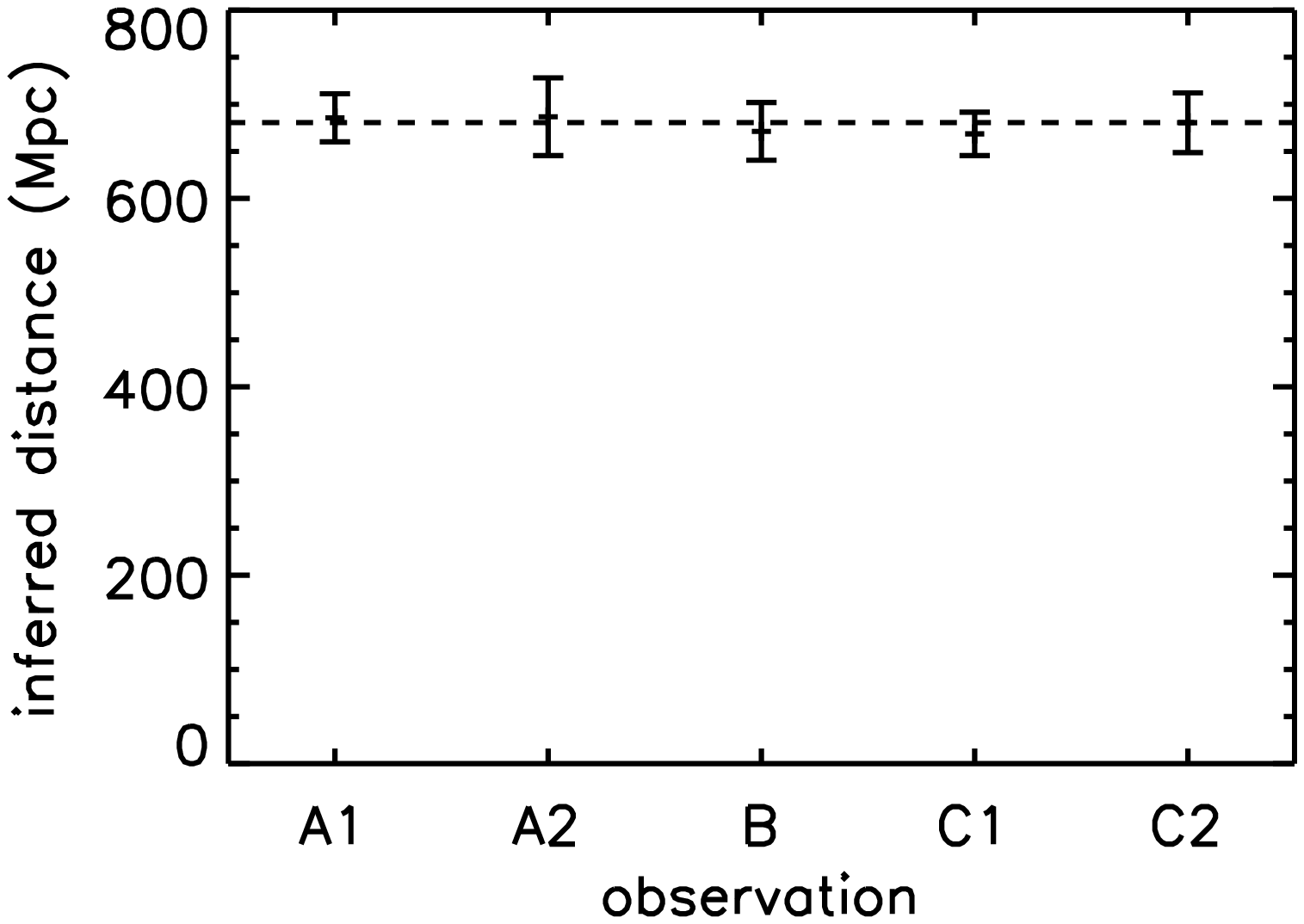}
\caption {Inferred distance for each trial observation (see
Table~\ref{trial_observations}) as estimated from 5 realizations of the
artificial observations.  The error bars are the sample standard
deviation of the 5 realizations, corresponding to the $1 \sigma$
uncertainties for a single realization.  \Label{model_distances}}
\end{figure}

\section{Discussion}
\label{discussion}

\subsection{Statistical Uncertainties}

The variation among realizations measures the sensitivity of the
inferred distance to statistical uncertainties in the X-ray and
temperature decrement observations.  By examining correlations between
the fractional errors in the distance and in the mass elongation, it is
possible to show that the uncertainty in the X-ray temperature makes
the dominant contribution to the uncertainty in the distance in the
case of clusters axisymmetric about line of sight \citep{Fox2001}.  In
the general case, however there was no obvious correlations between the
distance and the shape parameters, $Z_{ \rm mass }$ and $\svec$.

As we noted in \S\ref{lensing_uncertainties}, we have not attempted to
model the uncertainties in the gravitational lensing observations. 
Additional Monte Carlo tests with realistic treatment of these
uncertainties, and their effects on the fitting and de-projection of
the surface mass distribution, should be performed.

\subsection{Systematic Errors}

Besides the statistical measurement uncertainties, there are number of
systematic effects which could affect our determination of the cluster
distances and of $H_0$.  Many of these effects are also relevant to the
traditional method based on a spherical model.  For recent reviews of
the systematic uncertainties, see \citet{Carlstrom2001} and
\citet{MBM2002}.

We will focus primarily on those effects which are particular to our
method.  We first discuss the dependence of the inferred $H_0$ on the
magnitude of the observed quantities, and the resulting bias due to
either contamination of these observations or to errors in their
absolute calibration.  Then, we consider aspects of the state of the
intracluster gas which could violate the underlying physical
assumptions of our model.  Finally, we mention some caveats regarding
our assumption of ellipsoidal symmetry.

\subsubsection{Normalization Dependence}

The distance determination depends on the normalizations of the
measured temperature decrement, X-ray surface brightness, and X-ray
temperature.  Therefore, calibration errors or other systematic errors
in these observations will directly affect the value of $H_0$ inferred.
 In the traditional method,
\begin{equation}
H_0 \propto D^{-1}_{ \rm inferred } \propto { \Sigma_X T_e^2
\over ( \Delta T_{ \rm CMB } )^2 } , \label{traditional_systematics}
\end{equation}
assuming that the value for $H_0$ has been averaged over sufficiently
many clusters of random orientation to remove the effect of the
elongation.  For a single cluster, there would be an additional factor
of the inverse of the gas elongation, $Z$.  In our method, $H_0$ can,
in principle, be inferred from a single cluster by correcting for this
factor.  However, this correction introduces an additional dependence
on the ratio of the potential, $\phi$, determined from weak lensing
observations, to the X-ray temperature.  From
equations~(\ref{elongated_pressure}) and (\ref{predicted_lensing}), we
can show that the inferred gas elongation,
\begin{equation}
Z \propto { \Sigma_{ \rm cr } D_l \psi \over T_e }
\label{elongation_dependence}.
\end{equation}
Recall that the product $\Sigma_{ \rm cr } D_l$ is independent of the
distance to the lensing cluster, $D_l$, and depends only on the
cosmology and the redshifts of the cluster and the lensed sources. 
Combining equations~(\ref{traditional_systematics}) and
(\ref{elongation_dependence}), we see that for our method,
\begin{equation}
H_0 \propto D^{-1}_{ \rm inferred } \propto { \Sigma_{ \rm cr } D_l
\psi \Sigma_X T_e \over ( \Delta T_{ \rm CMB } )^2
}.\Label{new_systematics}
\end{equation}
Thus, our method and the traditional method have the same sensitivity
to errors in the normalization of the X-ray surface brightness or the
temperature decrement.  Possible sources of contamination of these
measurements, as well as uncertainties in their calibration, are
discussed by \citet{Carlstrom2001}.

Due to the fortuitous cancellation of one power of the electron
temperature between equations~(\ref{traditional_systematics}) and
(\ref{elongation_dependence}), we expect our method to be less
sensitive to the electron temperature than the traditional method. 
Nonetheless, the inferred value of $H_0$ will still be directly
proportional to the absolute calibration of the X-ray temperature, as
well as that of the surface brightness.

Our method introduces a new dependence on the calibration of the weak
lensing observations.  Our inferred $H_0$ is now proportional to the
normalization of the lensing potential, $\psi$, (or its second
derivatives, the convergence, $\kappa$, and the shear, $\gamma$) as
well as to
\begin{equation}
\Sigma_{ \rm cr } D_l = { c^2 \over 4 \pi G } { D_s \over D_{ ls } }.
\end{equation}
For measurements of weak shear from the shapes of background galaxies,
the normalization of the inferred shear depends on accurately
correcting for the seeing conditions.  Calculating $\Sigma_{ \rm cr }
D_l$ requires knowledge of the redshift distribution of the background
galaxies, and of the choice of cosmology ($\Omega_0$ and $\Lambda_0$)
which affect the shape of the redshift-distance relation.  Note that
all lensing measurements depend on $\Sigma_{ \rm cr }$.  Therefore, it
may be possible to eliminate the cosmology dependence if the redshift
distribution is itself inferred from lensing measurements.  Several
techniques have been suggested for doing so \citep{Smail94, Kneib94,
Kneib96III, BN95}.  

Our analytic models in \S\ref{degeneracy} demonstrated that the
elongation-distance degeneracy could be broken without resolving the
mass sheet degeneracy in the lensing observations.  However, while we
have proven this in principle, we have not yet demonstrated it in
practice.  In fact, our assumption in \S\ref{mass_model} that the
product $D_l \Sigma$ was known presumes that the mass sheet degeneracy
has been resolved.  This could be done by measuring the convergence. 
Alternatively, we can assume that the surface density falls to
negligible values far from the center of the cluster.  Given shear
observations over a wide enough field, this may be sufficient. 
Finally, the method might be reformulated to remove any dependence on
the mass sheet degeneracy.

\subsubsection{The Physical State of the Intracluster Gas}

In comparing $n_e^2$ inferred from the X-ray observations with $n_e$
inferred from the temperature decrement, both our method and the
traditional one assume that the X-ray emission and the scattering of
the microwave background photons is due to the same population of
electrons.  If the intracluster gas is not smooth but rather clumpy,
with $C \defeq \langle n_e^2 \rangle / \langle n_e \rangle^2 > 1$, then
the inferred distance to the cluster will be systematically
underestimated by a factor of $C$ \citep{BHA91}.  In addition, a
multiphase medium introduces additional degrees of freedom which may
make it impossible to constrain the elongation.  The gas in the cooling
flows observed in the centers of some clusters is in fact believed to
be multiphase.  In applying the our method to real clusters, it would
be advisable to remove any cooling flow region from the X-ray maps and
fit to the remaining data.

Our method of inferring the elongation depends on the assumption that
the intracluster gas is in hydrostatic equilibrium in the cluster
potential well.  A distinct advantage of the traditional method is that
it does not rely on this assumption.  The characteristic time for a
cluster to reach hydrostatic equilibrium is the sound wave crossing
time.  The high temperature of the intracluster gas makes this time
scale much shorter than the Hubble time \citep{Sarazin88}.  However,
clusters which have undergone recent mergers may not have had time to
fully relax.  Mergers tend to increase the X-ray luminosity of
clusters, so cluster samples selected by luminosity will tend to
include an enhanced fraction of mergers \citep{RSM97}.  Deviations from
hydrostatic equilibrium could introduce either random or systematic
errors in our elongation correction.

We have also implicitly assumed that thermal pressure is the dominant
source of support for the gas.  Other possible sources include
rotation, magnetic or turbulent pressure, or pressure from a population
of relativistic electrons and ions.  If these were significant, our
method would give incorrect results for the elongation and the
distance.  

It is essential to quantify these possible errors by applying the
method to a set of simulated clusters.

\subsubsection{Symmetry and Modeling Assumptions}

Real clusters will differ from the models we use to fit them, and this
may also introduce both random and systematic errors.  Typically,
measurements of $H_0$ from X-ray and SZ effect observations have used a
beta model to fit the gas density distribution.  Often, in the absence
of spatially resolved X-ray temperature measurements, an isothermal
temperature has also been assumed.  A number of authors have estimated
the systematic effects in the traditional method due to deviations from
these model assumptions, as well as that of spherical symmetry
\citep{Sulkanen99, ISS95, RSM97, Puy2000}.  We have tried to take
advantage of the improved spectral resolution of the latest X-ray
telescopes, and of anticipated improvements in observations of the SZ
effect, to relax these assumptions about the density and temperature
profiles of the gas, allowing them to the constrained only by the
observations and by the assumption of hydrostatic equilibrium.

Even so, de-projecting clusters still requires a fairly strong
assumption about symmetry.  We have assumed that the cluster mass
distribution is constant on families of similar, concentric, coaxial
ellipsoids.  Our hope is that such a model will be sufficient for
inferring to first order the deviation from spherical symmetry due to
elongation.  However, X-ray clusters often show evidence for
significant irregularity and substructure \citep[e.g.,][]{JF92,
Mohr95}.  If the deviations from our model are large enough, it may
fail to fit the observations, or, worse, give misleading answers.  For
example, our model would be unlikely to give correct results when
applied to a cluster which was actually two clusters along the same
line of sight (unless one of the two clusters dominated all three of
our observational probes, or unless the two shared a common envelope). 
Selection effects in the detection of gravitational lensing could bias
a sample of clusters toward such cases.  Even fairly regular clusters
will not have the perfect ellipsoidal symmetry we have assumed.  They
may be elongated but not ellipsoidal.  They may be roughly ellipsoidal,
but with smaller scale irregularities, or with axis ratios and
orientations which vary with radius.  Again, the best way to quantify
the resulting uncertainties, both random and systematic, is to apply
the method to a sample of simulated clusters, with realistic modeling
of the selection effects.

Mass in filaments and other correlated structure near clusters
contributes significantly to the total projected mass measured by
lensing, and to its scatter, though the magnitudes of both the bias and
the scatter depend on the particular mass estimator used
\citep{Metzler99, Metzler2001}.  By applying our method to numerically
simulated clusters, we will be able to tell how much of an effect this
has on the inferred distances.

\section{Conclusions}

\label{conclusions}

X-ray and SZ effect observations provide a geometric method of
measuring the distance to clusters of galaxies, independent of the
usual distance ladder calibrations.  This method can be used to measure
the redshift-distance relation and inferred the value of $H_0$. 
However, the method is subject to an exact degeneracy between the
distance and the elongation of the cluster along the line of sight,
relative to its size perpendicular to the line of sight.  This
degeneracy introduces significant scatter in the measured values of
$H_0$, requiring a large sample of clusters of random orientation for
an accurate determination.  To infer $\Omega_0$ and $\Lambda_0$ by the
redshift-distance relation would require an even larger sample. 
Furthermore, if the sample is biased towards particular orientations
due to selection effects, this would introduce a systematic error in
$H_0$.

The elongation-distance degeneracy can be broken if the intracluster
gas is in hydrostatic equilibrium, and if the projected mass density
can be inferred from gravitational lensing measurements of background
galaxies.  We have developed a specific method to do so, using models
with ellipsoidal mass distributions, and applied it to simple
axisymmetric ellipsoidal model clusters.  We recover the true shape and
distance to each cluster with statistical uncertainties of $4$ to $6
\%$, and no detectable systematic bias (with a $3 \sigma$ upper limit
of $4 \%$), using artificial X-ray and SZ effect observations with
sensitivity comparable to current observations.

Further work is necessary to determine the sensitivity of the method to
realistic uncertainties in the lensing observations.  Particular
attention should be paid to possible systematic uncertainties due to
the mass sheet degeneracy, and to the dependence of the critical
surface density on the redshift distribution of the lensed background
galaxies.  The systematic uncertainties in the calibration of the X-ray
surface brightness and temperature observations and the SZ effect
observations are also important.  

Finally, real clusters are not the perfectly regular, ellipsoidal
objects we have assumed in our models, nor are they necessarily in
perfect hydrostatic equilibrium.  It is essential to test the method by
applying it to numerically simulated clusters, which more closely
resemble real clusters, at least in their greater degree of messiness.

\acknowledgments

We would like to thank Abraham Loeb for suggesting combining X-ray and
lensing observations to infer the line of sight elongation of clusters,
and for his advice on an earlier incarnation of this project.  This
work was supported by a National Science Foundation Graduate Research
Fellowship for D. C. F..  Many thanks to Laura Grego and Chris Metzler
for useful discussions, and to Irwin Shapiro for pulling at a loose
thread in the preliminary results.  One of the authors (D. C. F.) would
like to acknowledge Dragon Systems, Inc., whose DragonDictate for
Windows and Dragon NaturallySpeaking software were essential to the
development of the code for this calculation and to the preparation of
this paper.

\appendix

\section{APPENDIX}
\label{projection_details}

Since the gas properties are constant on isopotentials, the projection
of a local gas quantity, $f ( \phi )$, is
\begin{equation}
\Sigma ( \thetavec ) = D \int d \zeta \, f \left ( \phi ( \thetavec ,
\zeta ) \right ).\Label{projection_equation}
\end{equation}
The projection operator is linear, and we already have a finite set of
$n$ lines of sight, $\thetavec_i$, so an obvious approach is to
discretize the local quantity, and calculate a projection matrix, $K$. 
We pick a finite grid in $\phi$, $\phi_0 < \phi_1 < \ldots < \phi_{ m -
1 } \defeq \phi_{ \rm max }$.  Then, the projection reduces to a
matrix-vector multiplication,
\begin{equation}
\Sigma ( \thetavec_i ) = \sum_j D K_{ ij } f_j
\end{equation}
where $f_j \defeq f ( \phi_j )$.  Similarly, the gradient with respect
to the gas parameters, $\avec$, is simply a matrix-matrix
multiplication,
\begin{equation}
\grad_{ \avec } \Sigma ( \thetavec_i ) = \sum_j DK_{ ij } \grad_{ \avec
} f_j
\end{equation}

To calculate the elements of the projection matrix, $K$, we replace $f
( \phi )$ in equation~(\ref{projection_equation}) with a linear
interpolation in $\phi$ between the grid points.  When $\phi < \phi_0$,
we let $f ( \phi ) \defeq f ( \phi_0 )$.  For each line of sight, we
cut off the integral at $\phi = \phi_{ \rm max }$.  The elements of
$K$, depend only on the potential, $\phi ( \xvec )$, the potential
grid, and the set of lines of sight, not on the local gas properties,
$f$ (or $\grad_{ \avec } f$).  Thus, we can use the same matrix for all
X-ray bands, and for all trials with a single assumed cluster shape.

The correct choice of the cutoff, $\phi_{ \rm max }$, is rather
delicate.  Because it is proportional to the electron density squared,
the X-ray emission along a given line of sight is dominated by the
contribution from isopotentials near the minimum of $\phi ( \thetavec ,
\zeta )$ with respect to $\zeta$.  If $\phi_{ \rm max }$ is too large,
none of the lines of sight will have a minimum potential comparable to
it.  In that case, the gas properties at that isopotential will be
poorly constrained, and minimizing $\chi^2$ may lead to an unphysical
temperature profile.    On the other hand, if $\phi_{ \rm max }$ is too
small, the cluster emission will be artificially cut off at some radius
within the field of view.  This radius may not be the same for both the
model cluster and the true cluster, which will lead to systematic
problems in the fitting.  In addition, fixing $\phi_{ \rm max }$ to a
single value for all models tends to bias the shape fitting.  More
elongated de-projections have shallower potential wells, so a fixed
grid in $\phi$ effectively reduces the angular resolution of these
models, making their $\chi^2$ values artificially large.

To avoid all these difficulties, we want to choose $\phi_{ \rm max }$
for each model shape so as to ensure that the outermost lines of sight
in the field view are tangent to $\phi = \phi_{ \rm max }$.  In the
axisymmetric case, this simply means setting $\phi_{ \rm max }$ equal
to the minimum of the potential along a line of sight with the maximum
radius.  In the general case, we evaluate the minimum of the potential
along the lines of sight at each corner and in the center of each edge
of the field of view, and choose $\phi_{ \rm max }$ equal to the
largest such minimum.  We use a uniform grid in the potential with 20
elements between $0$ and $\phi_{ \rm max }$.


\clearpage
\newpage

\end{document}